\documentclass[aps,twocolumn,prl,groupaddress,floatfix]{revtex4}

\usepackage{graphicx}
\usepackage{dcolumn}
\usepackage{bm}

\begin{document}

\title{Many-body orbital paramagnetism in doped graphene sheets}
\author{A. Principi}
\affiliation{NEST, Istituto Nanoscienze-CNR and Scuola Normale Superiore, I-56126 Pisa, Italy}
\author{Marco Polini}
\email{m.polini@sns.it} \homepage{http://qti.sns.it/}
\affiliation{NEST, Istituto Nanoscienze-CNR and Scuola Normale Superiore, I-56126 Pisa, Italy}
\author{G. Vignale}
\affiliation{Department of Physics and Astronomy, University of Missouri, Columbia, Missouri 65211, USA}
\author{M.I. Katsnelson}
\affiliation{Radboud University Nijmegen, Institute for Molecules and Materials, NL-6525 AJ Nijmegen, The
Netherlands}

\begin{abstract}
The orbital magnetic susceptibility (OMS) of a gas of {\it
noninteracting} massless Dirac fermions is {\it zero} when the
Fermi energy is away from the Dirac point. Making use of diagrammatic
perturbation theory, we calculate exactly the OMS of massless Dirac
fermions to first order in the Coulomb interaction demonstrating
that it is finite and {\it positive}. Doped graphene sheets are
thus unique systems in which the OMS is completely controlled by
many-body effects.
\end{abstract}

\maketitle

{\it Introduction.}--- The diamagnetic properties of carbon allotropes such as diamond
and graphite have always attracted a great deal of interest. Early
on, it was shown experimentally that the diamagnetic susceptibility
of graphite is large and strongly
anisotropic~\cite{krishnan_nature_1934}. These studies stimulated
an intense theoretical activity~\cite{mcclure_pr_1956}. The recent
isolation of graphene (see Ref.~\onlinecite{reviews} for reviews)
has revitalized an interest in the magnetic properties of
carbon-based materials. Graphene is a truly two-dimensional (2D)
system composed of carbon atoms tightly packed in a honeycomb
lattice. States near the Fermi energy of a graphene sheet are
described by a massless Dirac equation which has chiral states in
which the honeycomb-sublattice pseudospin is aligned either
parallel to or opposite to momentum~\cite{reviews}. McClure was
the first one to realize that the Landau quantization for 2D
massless Dirac fermions (MDFs) in a magnetic field perpendicular
to the layer is very special (existence of a zero-energy Landau level) and
to conjecture that this is the cause of the large diamagnetic
susceptibility of graphite~\cite{mcclure_pr_1956,reviews}.

More precisely, McClure showed that the orbital magnetic
susceptibility (OMS) of a noninteracting gas of 2D MDFs at a
finite temperature $T$ is given by~\cite{mcclure_pr_1956,safran_prb_1979,koshino_prb_2007,principi_prb_2009,literature}
\begin{eqnarray}\label{eq:orb_T}
\chi^{(0)}_{\rm orb} &=& -\frac{g_{\rm s} g_{\rm v}}{24 \pi} \frac{e^2 v^2}{c^2} 
\frac{1}{k_{\rm B} T \cosh^2{[\mu/(2 k_{\rm B} T)]}}  \nonumber\\
&\stackrel{T\to 0}{=}& - \frac{g_{\rm s}g_{\rm v}}{6\pi} \frac{e^2 v^2}{c^2} \delta(\varepsilon_{\rm F})~,
\end{eqnarray}
where $g_{\rm s} = g_{\rm v} = 2$ are spin and valley degeneracies
factors, $v$ is the Fermi velocity (which is independent on
carrier density), $c$ is the speed of light, $\mu$ is the chemical
potential, and $\varepsilon_{\rm F} = \mu (T=0)$ is the Fermi
energy. The zero-temperature OMS of graphene is thus infinite in
the undoped limit, {\it i.e.} in the limit of zero carrier
density. The second line of Eq.~(\ref{eq:orb_T}) encodes another
astonishing result~\cite{koshino_prb_2007,principi_prb_2009}. The
OMS of graphene is exactly {\it zero} (at $T=0$) if the Fermi energy is away
from the Dirac point, {\it i.e.} if the system is doped. This
situation seems to be really unique.

Eq.~(\ref{eq:orb_T}) (and all the other studies~\cite{literature}
of the orbital properties of MDFs we are aware of) heavily relies
on a single-particle picture. The fundamental question we address
in this Letter is the impact of electron-electron interactions on
the OMS of a doped graphene sheet. Short-range repulsive
interactions in a neutral Fermi gas~\cite{books,jo_science_2009},
for example, or Coulomb interactions in an ordinary parabolic-band
electron gas~\cite{Giuliani_and_Vignale} enhance the paramagnetic
nature of the spin response (eventually driving the system toward
a ferromagnetic
instability~\cite{jo_science_2009,Giuliani_and_Vignale}) but are
typically never strong enough to switch the sign of the orbital
response from diamagnetic to paramagnetic. Orbital paramagnetism
(OP), although possible in principle, is indeed a rare phenomenon
in metals and semiconductors. For example, it has been shown that
a 2D electron gas in a periodic potential exhibits OP when the
Fermi level is sufficiently close to a saddle point of the band
structure~\cite{vignale_prl_1991}. Electrons in the proximity to a
superconductor have been shown~\cite{bruder_prl_1998} to exhibit
OP. Using diagrammatic perturbation theory up to first order in
the Coulomb interaction we will demonstrate that the OMS of an
interacting gas of 2D MDFs is finite and positive.
Weakly-interacting doped graphene sheets thus represent unique
systems with a paramagnetic orbital response of purely many-body
origin.

\noindent {\it MDF model Hamiltonian and linear-response theory.}--- The (single-channel) Hamiltonian of a 2D gas of MDFs in the eigenstate representation is ($\hbar =1$)
\begin{equation}\label{eq:MDFhamiltonian}
{\hat {\cal H}} = \sum_{{\bm k}, \lambda} \varepsilon_{{\bm k}, \lambda} {\hat c}^\dagger_{{\bm k}, \lambda}{\hat c}_{{\bm k}, \lambda} +\frac{1}{2S}\sum_{{\bm q} \neq 0} v_q {\hat \rho}_{\bm q} {\hat \rho}_{-{\bm q}}~,
\end{equation}
where $\varepsilon_{{\bm k}, \lambda} = \lambda v k$ ($\lambda = \pm$) are band energies, $S$ is the area of the system,
$
{\hat \rho}_{\bm q} = \sum_{{\bm k}, \lambda, \mu}
M_{\lambda\mu}({\bm k},{\bm q})
{\hat c}^\dagger_{{\bm k}-{\bm q}/2, \lambda}
{\hat c}_{{\bm k}+{\bm q}/2, \mu}
$
is the density operator, $v_q = 2\pi e^2/(\epsilon q)$ is the 2D Fourier transform of the Coulomb potential, $\epsilon$ is an average dielectric constant, and $M_{\lambda\mu}({\bm k},{\bm q})$ are matrix elements which can be found, for example, in Ref.~\onlinecite{principi_prb_2009}. The many-body properties of doped graphene sheets
depend~\cite{reviews,barlas_prl_2007} on the dimensionless coupling constant (restoring $\hbar$ for a moment) $\alpha_{\rm ee}= e^2/(\epsilon \hbar v)$, which
can be tuned experimentally by changing the dielectric environment
surrounding the graphene flake~\cite{dielectric_environment}. The
model (\ref{eq:MDFhamiltonian}) requires the introduction of an ultraviolet cut-off
$k_{\rm max}$~\cite{barlas_prl_2007}  on the ${\bm k}$-sums.  
This must be done with great care, since,  as discussed in
Ref.~\onlinecite{principi_prb_2009}, the presence of
$k_{\rm max}$ breaks gauge invariance, which must be restored in the calculations.

\begin{figure}
\begin{center}
\includegraphics[width=1.0\linewidth]{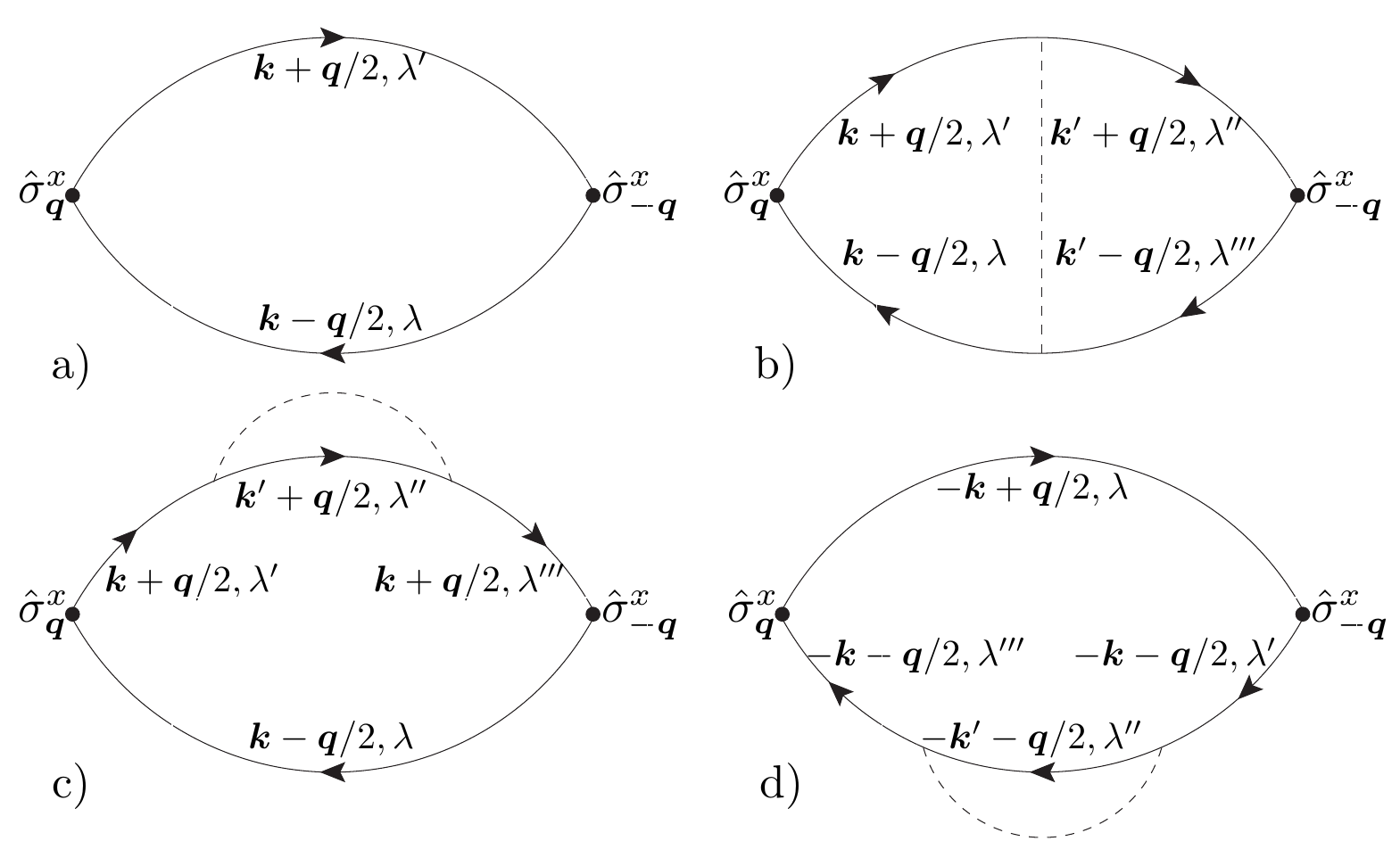}
\caption{ Diagrams for the orbital magnetic susceptibility.  a) The noninteracting bubble. Panel b) The first-order vertex correction.
Panels c) and d) The two first-order self-energy diagrams. The wave-vector labeling of the propagators ensures time-reversal symmetry and thus guarantees that the two self-energy diagrams give an identical contribution.\label{fig:one}}
\end{center}
\end{figure}

The OMS of the system described by ${\hat {\cal H}}$, $\chi_{\rm orb}$,
can be calculated from~\cite{Giuliani_and_Vignale}
\begin{equation}\label{eq:LRT}
\chi_{\rm orb} = -\frac{e^2}{c^2} \lim_{q \to 0} \frac{\chi(q)}{q^2}~,
\end{equation}
where $\chi(q)$ is a short-hand notation for the static transverse
current response function. The bare current-density
operator ${\hat {\bm j}}_{\bm q}$ for MDFs is proportional to the
pseudospin-density operator~\cite{reviews}: ${\hat {\bm j}}_{\bm
q} = v {\hat {\bm \sigma}}_{\bm q}$. In what follows we will
calculate $\chi(q)$ up to order $q^2$ in the long-wavelength $q\to
0$ limit using many-body diagrammatic perturbation theory.

\noindent {\it Diagrammatic perturbation theory.}--- Fig. 1 shows the four diagrams that contribute to the transverse (${\bm q} = q {\hat {\bm y}}$) pseudospin response function
$\chi(q)$ to first order in the electron-electron interactions.
Solid lines are noninteracting Green's functions in the eigenstate representation,
$
G_\lambda ({\bm k}, \omega)  = [\omega - \varepsilon_{{\bm k}, \lambda} + i \eta_{{\bm k}, \lambda}]^{-1}
$
where  $\eta_{{\bm k}, \lambda} = \eta \lambda~{\rm sgn}(k -
k_{{\rm F}, \lambda})$, with $\eta = 0^+$.  Dashed lines are electron-electron interactions.
Here $k_{{\rm F}, +} = k_{\rm F} = \sqrt{ 4 \pi n/(g_{\rm s} g_{\rm v})}$ is the Fermi 
wave number corresponding to an electron concentration $n$~\cite{phsymmetry} and $k_{{\rm F}, -} = k_{\rm max}$.    

In the static limit the zeroth-order bare-bubble diagram in panel a) gives $\chi^{(0)}(q \to
0) = a_0 + a_2 q^2$, with $a_0 = - k_{\rm max}/(4\pi v)$ and $a_2
=0$. As discussed in Ref.~\onlinecite{principi_prb_2009}, the term
${\cal O}(q^0)$ has to be subtracted away to restore
gauge-invariance. The fact that $a_2 =0$ originates from a perfect
cancellation of intra- and inter-band contributions. After the
{\it ad~hoc} regularization $a_0 \equiv 0$, we find, in agreement with the second line of Eq.~(\ref{eq:orb_T}), that the
OMS of the doped noninteracting system is zero: $\chi^{(0)}_{\rm
orb} =0$.

Panels b), c), and d) in Fig.~\ref{fig:one} show the remaining first-order diagrams:  
panel b) is the so-called ``vertex correction",
while panels c) and d) contain two ``self-energy" insertions. The
evaluation of these diagrams up to order $q^2$ is lengthy 
and will be presented elsewhere. To
achieve analytical progress it turns out to be particularly
useful~\cite{borghi_ssc_2009}  to decompose the isotropic interaction $v_{{\bm k}-{\bm k}'}$ in
Eq.~(\ref{eq:MDFhamiltonian}) in angular momentum components as
$v_{{\bm k}-{\bm k}'} = \sum_{m} V_m(k,k') e^{i m (\varphi_{\bm
k}-\varphi_{{\bm k}'})}$ with $\varphi_{\bm k} = {\hat {\bm k}} \cdot {\hat {\bm x}}$ and the pseudopotentials $V_m(k,k')$ defined in Eq.~(10) of Ref.~\onlinecite{borghi_ssc_2009}. 
In what follows we will introduce dimensionless variables: wave vectors
will all be measured in units of $k_{\rm F}$, the pseudopotentials $V_{m}(k,k')$ in units of $2\pi e^2/(\epsilon
k_{\rm F})$, and the response function $\chi(q)$ in units of the
single-channel MDF density-of-states $\nu(\varepsilon_{\rm F}) =
\varepsilon_{\rm F}/(2\pi v^2)$, $\varepsilon_{\rm F} = v k_{\rm
F}$. The ultraviolet cut-off $k_{\rm max}$ measured in units of $k_{\rm F}$ will be denoted by $\Lambda = k_{\rm
max}/k_{\rm F}$.

Modulo terms ${\cal O}(q^0)$, which, as mentioned above, must be
removed by hand to restore gauge
invariance~\cite{principi_prb_2009}, the static transverse
response function $\chi(q)$ to first order in $\alpha_{\rm ee}$
reads
\begin{eqnarray}\label{eq:transverseresponsefinal}
\chi(q) = g_{\rm s} g_{\rm v} \alpha_{\rm ee} \nu(\varepsilon_{\rm F}) q^2 (\Xi_1 + \Xi_2 + \Xi_3)~,
\end{eqnarray}
where the three dimensionless coefficients $\Xi_n$ are given by
the sum of vertex $\Xi^{({\rm VC})}_n$ and self-energy $\Xi^{({\rm
SE})}_n$ contributions, $\Xi_n \equiv \Xi^{({\rm VC})}_n +
\Xi^{({\rm SE})}_n$:
\begin{widetext}
\begin{eqnarray}
\Xi^{({\rm VC})}_1 &=& \frac{1}{128}\int_1^\Lambda \frac{dk}{k^2} \int_1^k \frac{dk'}{k'^2} (k^2 + k'^2) \Big[ 5 V_0(k,k') + V_2(k,k') \Big] \equiv \int_{1}^\Lambda dk~f_1^{({\rm VC})}(k)~, \label{eq:vc_f_1} \\
\Xi^{({\rm VC})}_2 &=& \frac{1}{384} \int_1^\Lambda \frac{dk}{k^2} \Big\{ 15(1 - k^2) V_0(k,1) -3 (1 + k^2) V_2(k,1) + 2 k^2 \partial_{k'}[8  V_0 (k,k') + V_2(k,k') ]_{k' = 1} \nonumber\\
&-& k^2 \partial_{k'}^2 [ V_0(k,k') - V_2 (k,k')]_{k' = 1} \Big\}
\equiv \int_{1}^\Lambda dk~f_2^{({\rm VC})}(k)~, \label{eq:vc_f_2} \\
\Xi^{({\rm VC})}_3 &=& -\frac{1}{384} \Big\{ 3 [21 V_0(1,1) - 6 V_1(1,1) - V_2(1,1)] - 2 \partial_{k'} [8 V_0(k,k') + 7 V_1(k,k') - V_2(k,k')]_{k = k' = 1} \nonumber\\
&+& \partial_{k'}^2 [V_0(k,k') + 2 V_1(k,k') + V_2(k,k')]_{k = k' = 1} \Big\}
\equiv 2 \int_{0}^\pi d\theta~f_3^{({\rm VC})}(\theta)~,\label{eq:vc_f_3} \\
\Xi^{({\rm SE})}_1 &=& \frac{1}{128} \int_1^\Lambda dk \int_1^k dk' \left[ 5\frac{k^2 + k'^2}{k^2 k'^2} V_0(k,k') - 3\frac{5 k^4 + k^2 k'^2 + 5 k'^4}{k^3 k'^3} V_1(k,k') + \frac{k^2 + k'^2}{k^2 k'^2} V_2(k,k') + \frac{9}{k k'}V_3(k,k')\right]~, \nonumber \\
\label{eq:se_f_1}\\
\Xi^{({\rm SE})}_2 &=& \frac{1}{768} \int_1^\Lambda \frac{dk}{k^2} \Big\{ 3(11 - 10 k^2) V_0(k,1) + 3 k (36 k^2 - 5) V_1(k,1) - 3 (10 k^2 + 1) V_2(k,1) - 81 k V_3(k,1) \nonumber\\
&+& 30 V_4(k,1) - \partial_{k'}[7 k^2  V_0(k,k') + 9 k (2 k^2 + 1) V_1(k,k') - 7 k^2 V_2(k,k') + 9 k V_3 (k, k')]_{k'=1} \nonumber\\
&+& k^2 \partial_{k'}^2 [V_0(k,k') - V_2(k,k')]_{k'=1} \Big\}~, \label{eq:se_f_2}
\end{eqnarray}
and
\begin{eqnarray}
\Xi^{({\rm SE})}_3 &=& -\frac{1}{768} \Big\{ 33 V_0(1,1) - 72 V_1(1,1) - 27 V_2(1,1) - 48 V_3(1,1) + 30 V_4(1,1) \nonumber\\
&+& \partial_{k'} [7 V_0(k,k') + 23 V_1(k,k') + 5 V_2(k,k') + 5 V_3(k,k') + 16 V_4(k,k')]_{k=k'=1} \nonumber\\
&-& \partial_{k'}^2 [4 V_0(k,k') + 8 V_1(k,k') + 2 V_2(k,k') - 4 V_3(k,k') - 2 V_4(k,k')]_{k=k'=1} \Big\}
\equiv 2 \int_{0}^\pi d\theta~f_3^{({\rm SE})}(\theta)~.\label{eq:se_f_3}
\end{eqnarray}
\end{widetext}
It is easy to prove that $f_1^{({\rm VC})}(k \to \infty) \to 5/(128 k) +
5 (18C -7)/(576 \pi k^2) + {\cal O}(k^{-3})$ and that $f_2^{({\rm
VC})}(k \to \infty) \to -5/(128 k) + {\cal O}(k^{-3})$. Here $C
\simeq 0.916$ is Catalan's constant. This implies that the
sum $f_1^{({\rm VC})}(k) + f_2^{({\rm VC})}(k)$ decays like
$k^{-2}$ for large $k$: we can thus take the limit $\Lambda \to
\infty$ in Eqs.~(\ref{eq:vc_f_1}) and~(\ref{eq:vc_f_2}) finding a
finite, cut-off independent result for the sum $\Xi^{({\rm VC})}_1
+ \Xi^{({\rm VC})}_2$. This property is identically shared by the
sum $\Xi^{({\rm SE})}_1 + \Xi^{({\rm SE})}_2$. Thus vertex and
self-energy contributions are separately convergent in the
ultraviolet limit. It is however crucial to take into account both
contributions to have a finite result. Indeed, the quantities
$\Xi^{({\rm VC})}_3$ and $\Xi^{({\rm SE})}_3$ are separately
strongly divergent: for the Coulomb potential we find $f_3^{({\rm
VC})}(\theta \to 0) = - f_3^{({\rm SE})}(\theta \to 0) = 1/(192
\pi \theta^3)$ at small angles. The subleading terms in the Taylor
expansions of $f_3^{({\rm VC})}(\theta)$ and $f_3^{({\rm
SE})}(\theta)$ are also singular ($\propto 1/\theta$) but these
singularities do not cancel out upon summing $f_3^{({\rm
VC})}(\theta)$ with $f_3^{({\rm SE})}(\theta)$ and are an artifact
of first-order perturbation theory, which misses screening. These
pathologies are commonly cured~\cite{Giuliani_and_Vignale}
by using a statically-screened Thomas-Fermi interaction:
(restoring units for a moment) $v_q = 2\pi e^2/[\epsilon(q +
q_{\rm TF})]$,  $q_{\rm TF} = g_{\rm s} g_{\rm v} \alpha_{\rm ee}
k_{\rm F}$ being the Thomas-Fermi screening wave
number~\cite{TF}.

\begin{figure}
\begin{center}
\includegraphics[width=1.0\linewidth]{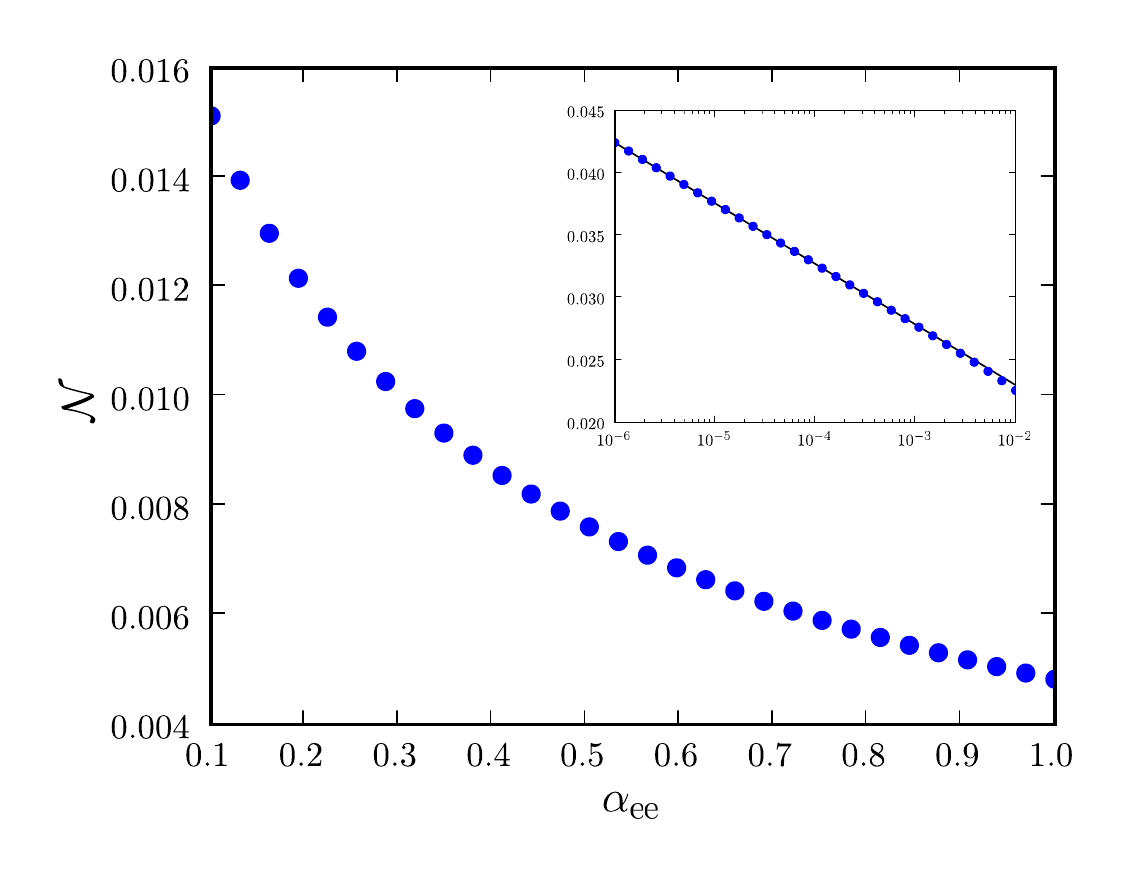}
\caption{(Color online) The dimensionless quantity ${\cal N} = - (\Xi_1 + \Xi_2 + \Xi_3)/(2\pi)$ as a function of the fine-structure constant $\alpha_{\rm ee}$.
Note that ${\cal N}(\alpha_{\rm ee})>0$ and that it depends weakly on $\alpha_{\rm ee}$. Inset: a zoom for very small $\alpha_{\rm ee}$. The horizontal axis is in {\it logarithmic scale}. The solid line represents the analytical weak-coupling result, ${\cal N}(\alpha_{\rm ee} \to 0) \to -\ln(\alpha_{\rm ee})/(48 \pi^2)$.\label{fig:two}}
\end{center}
\end{figure}

Substituting Eq.~(\ref{eq:transverseresponsefinal}) into
(\ref{eq:LRT}) we finally find that the OMS is given by
\begin{equation}\label{eq:final}
\chi_{\rm orb} = g_{\rm s} g_{\rm v} \frac{e^2 v^2}{c^2} \frac{\alpha_{\rm ee}~{\cal N}(\alpha_{\rm ee})}{\varepsilon_{\rm F}}~,
\end{equation}
where we have defined ${\cal N}(\alpha_{\rm ee}) = - (\Xi_1 +
\Xi_2 + \Xi_3)/(2\pi)$. Note that the final result
(\ref{eq:final}) is formally beyond the first order in
$\alpha_{\rm ee}$ since Thomas-Fermi screening introduces
non-linear dependencies on $\alpha_{\rm ee}$. The quantity ${\cal
N}$ is presented in Fig.~\ref{fig:two}.
One can see, first of all, that the function ${\cal N}(\alpha_{\rm
ee})$ is {\it positive}, which corresponds to OP, and depends
weakly on $\alpha_{\rm ee}$. It is easy to show that ${\cal
N}(\alpha_{\rm ee} \to 0) \to -\ln(\alpha_{\rm ee})/(48
\pi^2)$, implying that $\chi_{\rm orb} \propto - \alpha_{\rm ee}
\ln{(\alpha_{\rm ee})}$ in the weak-coupling limit~\cite{dynamical}. We remark that
the paramagnetic nature of the orbital response, {\it i.e.}
$\chi_{\rm orb} >0$, is stable with respect to changes in the
range of inter-particle interactions. Using the Hamiltonian
(\ref{eq:MDFhamiltonian}) with contact repulsive interactions of
strength $v_0$ ({\it i.e.} $v_q =v_0>0$), we find the analytical
expression
\begin{equation}
\chi_{\rm orb} = g_{\rm s} g_{\rm v} \frac{e^2}{c^2} \frac{13}{256 \pi^2} v_0~.
\end{equation}

\noindent {\it Discussion and conclusions.}--- Our findings can be
tested experimentally in a variety of ways. The most direct one is
to measure the thermodynamic magnetization as a function of an
applied magnetic field. Ingenious setups for these type of
measurements have already been successfully applied to various
carbon structures~\cite{thermodynamics_carbon} and to conventional 2D electron gases~\cite{thermodynamics_silicon}.
Spin and orbital responses can be distinguished by applying a
tilted magnetic field~\cite{fang_pr_1968}. To enhance the magnetic
response one can resort to graphene laminate~\cite{blake_nano_2008} (actually,
the first magnetic measurements in this system have already been done~\cite{magn_meas}). Another possibility is to use 
macroscopically large graphene films from carbon on copper foil, as demonstrated in Ref.~\onlinecite{bae_arxiv_2009}.

Interestingly, OP in doped graphene can be probed also by acoustic
measurements. Indeed, mechanical deformations of graphene are known to produce a pseudomagnetic gauge
field~\cite{guinea_natphys_2009}. The induced vector potential
${\bm A}({\bm r})$ is proportional to the deformation tensor
$u_{ij}({\bm r})$ and the pseudomagnetic field is given by the
usual relation ${\bm B}_{\rm S}({\bm r}) =\nabla_{\bm r} \times
{\bm A}({\bm r})$. Using the basic equations of the theory of
elasticity it is easy to prove that the coupling between
deformations and electronic degrees of freedom, which occurs {\it
via} ${\bm A}({\bm r})$, leads to a renormalization of the shear
modulus $\mu_{\rm s}  \to \mu_{\rm s}(q) = \mu_{\rm s} -
g^2_2\chi_{\rm orb} q^2/e^2$ and thus of the transverse sound
velocity $\omega^2(q) =\mu_{\rm s}(q) q^2/\rho$, where $g_2$ is a
coupling constant~\cite{gibertini_prb_2010} and $\rho$ is the mass
density. The positive sign of $\chi_{\rm orb}$ thus implies a
softening of the sound velocity with increasing $q$ whereas
diamagnetic response would result in the opposite behavior.

In summary, we have shown that doped graphene sheets have a very
intriguing orbital magnetic response. If electron-electron
interactions are neglected, the OMS is identically zero. When
electron-electron interactions are taken into account the OMS
turns out to be finite. To the best of our knowledge, this is the
first system we are aware of in which many-body effects control
completely the orbital response. The sign of the OMS cannot be
predicted {\it a~priori}. To first order in Coulomb interactions
we have shown that it is {\it positive}. Weakly-interacting doped
graphene sheets are thus many-body orbital paramagnets.

\noindent{\it Acknowledgements.}---  We acknowledge financial support by the 2009/2010 
CNR-CSIC scientific cooperation project (M.P.), by the NSF grant No. DMR-0705460 (G.V.), and by FOM, the Netherlands (M.I.K.).
We are grateful to A. Geim and I. Grigorieva for discussions on Ref.~\onlinecite{magn_meas} prior to publication.


\begin{thebibliography}{77}
\bibitem{phsymmetry}
    The model's particle-hole symmetry guarantees that electron-doped and hole-doped systems have identical OMS.
\end{thebibliography}
\end{document}